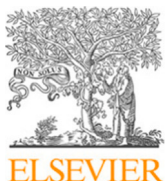
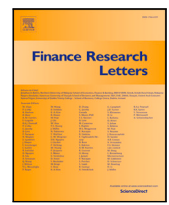

Contents lists available at ScienceDirect

## Finance Research Letters

journal homepage: www.elsevier.com/locate/frl

# Sentiment trading with large language models

Kemal Kirtac [a],[*], Guido Germano [a],[b]

[a] *Department of Computer Science, University College London, 66–72 Gower Street, London WC1E 6EA, United Kingdom*
[b] *Systemic Risk Centre, London School of Economics and Political Science, Houghton Street, London WC2A 2AE, United Kingdom*

**ABSTRACT**

We analyse the performance of the large language models (LLMs) OPT, BERT, and FinBERT, alongside the traditional Loughran-McDonald dictionary, in the sentiment analysis of 965,375 U.S. financial news articles from 2010 to 2023. Our findings reveal that the GPT-3-based OPT model significantly outperforms the others, predicting stock market returns with an accuracy of 74.4%. A long-short strategy based on OPT, accounting for 10 basis points (bps) in transaction costs, yields an exceptional Sharpe ratio of 3.05. From August 2021 to July 2023, this strategy produces an impressive 355% gain, outperforming other strategies and traditional market portfolios. This underscores the transformative potential of LLMs in financial market prediction and portfolio management and the necessity of employing sophisticated language models to develop effective investment strategies based on news sentiment.

## 1. Introduction

Many texts aim to understand and predict economic and financial events. In recent decades, the financial literature has turned to various sources of text data, such as financial news articles, regulatory filings, and social media posts, to extract valuable insights. However, the integration of text mining into financial models is still in its early stages. So far, most research has only explored a small portion of text data related to financial markets, often focusing on a single data source. Typically, these studies use straightforward methods, such as determining sentiment scores from dictionaries (Loughran and Mcdonald, 2011; Malo et al., 2014). A common finance-specific dictionary used in sentiment analysis is the Loughran-McDonald master dictionary (Loughran and McDonald, 2022).

Utilizing economic and financial information in text data poses difficulties. Unlike numerical data, text data lacks consistent structure, making it more complex to handle and interpret. Language itself is inherently intricate. Therefore, sophisticated models are necessary to uncover concealed insights from text. However, these advanced models can pose challenges, particularly for researchers who lack the technical expertise, computational resources, or sufficient funding. Given this context, it becomes apparent that current efforts in textual analysis within finance and economics are only beginning to explore the possibilities. There is a significant untapped potential in textual data that can be harnessed to gain a deeper understanding of asset markets. As challenges in text analysis emerge, they signal an exciting avenue for further research. It is expected that, in the future, economists will expand their text datasets and refine their techniques to extract valuable insights (Acemoglu et al., 2022).

The primary goal of this paper is to get deeper into this exploration. We aim to create more refined representations of news text using large language models (LLMs) and leverage these representations to develop models for predicting stock returns. To appreciate the value of LLMs, it is essential to understand the current landscape of financial text mining. Most methods in current use rely on supervised machine learning and are tailored for specific objectives, such as predicting returns, volatility, or macroeconomic variables (Jegadeesh and Wu, 2013; Baker et al., 2016; Manela and Moreira, 2017). These methods typically involve two main steps: converting text into a numerical representation and then modelling it. In many cases, researchers opt for the dictionary approach, which transforms each document into a vector that accounts for term frequency (Loughran and Mcdonald, 2011; Malo et al., 2014). Sometimes, this representation is further refined to create summarized scores that represent the text according to a domain-specific word dictionary (Tetlock, 2007). The outcome of the initial step is a matrix, which is then input into the econometric model in the subsequent step to describe various economic or financial phenomena. Studies such as Baker and Wurgler (2006), Lemmon and Ni (2014) and Shapiro et al. (2022) have highlighted the impact of investor sentiment on asset prices leveraging macro- or market-level sentiment indicators. However, these contributions do not deal with sentiment dynamics at the individual stock level, which our research aims to do by employing LLMs for a granular sentiment analysis. This approach not only enhances our understanding of sentiment's role in financial markets, but also introduces a novel dimension to sentiment analysis by focusing on the predictive power of LLM-derived sentiment scores for individual stock performances.

Current methods for representing financial text data have inherent limitations. Starting with the dictionary method, it oversimplifies text by solely considering term frequency and overlooking important aspects like word order and contextual relationships between terms (Malo et al., 2014). Additionally, the high dimensionality of a dictionary can lead to statistical inefficiencies, requiring many parameters in the subsequent modelling phase, even when many terms might be irrelevant. Dimensionality reduction techniques like latent Dirichlet allocation (LDA) can help, but are still based on a dictionary and do not address the fundamental issue of information loss (Bybee et al., 2019). Moreover, these reduced representations are specific to a text source in a particular domain, even though incorporating multiple text sources might provide better insights (Devlin et al., 2019).

A LLM can play a crucial role in text analysis. A LLM is trained on a wide variety of texts that cover different topics (Devlin et al., 2019). Creating an LLM involves a specialized team that builds a versatile language model. This model is trained on a vast amount of text, including entire books, Wikipedia articles, and more. Once developed, LLMs are made accessible for broader research purposes (Hugging Face, 2023).

In our research, we utilize two distinct LLMs for our analysis: Bidirectional Encoder Representations from Transformers (BERT) developed by Google (Devlin et al., 2019) and Open Pre-trained Transformers (OPT) developed by Meta (Zhang et al., 2022). State-of-the-art LLMs have consistently outperformed in various natural language processing (NLP) tasks, primarily due to their extensive scale. They are often available pre-trained on platforms like Hugging Face (Hugging Face, 2023). Hugging Face is a leading open-source repository that offers a plethora of advanced NLP and AI models, making state-of-the-art machine learning techniques readily available and deployable for academic and research purposes. We constructed our analytical models by leveraging the Hugging Face transformers library (Wolf et al., 2020) and utilizing Python.

We chose BERT for its innovation in language understanding, which is critical for analysing complex financial texts. BERT's ability to analyse the context of text from all sides improves how we interpret financial documents (Devlin et al., 2019). We would have liked to use also GPT-4 as a LLM in our research. However, we could not use the most advanced GPT models because they are not publicly available. Instead, we used OPT as a substitute. OPT is similar to the GPT models, but is open to the public (Radford et al., 2018). We used smaller versions of OPT due to our limited computer resources. For our analysis we chose BERT and OPT models with 345 million and 2.7 billion parameters (Hugging Face, 2023). Our choice of OPT and BERT provides us with powerful LLM tools to better understand financial markets and predict their movements. Furthermore, our analysis includes FinBERT, a specialized variant of the BERT LLM, specifically pre-trained for financial contexts. FinBERT is an example of how open-source BERT models can be adapted for distinct tasks in finance. Huang et al. (2023) demonstrated this adaptability by fine-tuning a BERT model for classification tasks using the Financial PhraseBank dataset, initially compiled by Malo et al. (2014). Last, we employed also the Loughran and McDonald dictionary (Loughran and Mcdonald, 2011; Loughran and McDonald, 2022) to facilitate a comparative assessment across different modelling approaches, i.e. FinBERT and our fine-tuned versions of BERT and OPT.

Our analysis is a two-step process: first, we turn text into numbers, and then we use those numbers to model economic patterns. In the first part, we choose how to represent text data numerically for the model in the second part. A popular method is the dictionary approach, which represents each text as a list of all the words and how often they appear. Sometimes, this basic list is all we use (Loughran and Mcdonald, 2011; Jegadeesh and Wu, 2013). Other times, we use the advanced LLM capabilities to analyse the news. The result of this first step is a table of numbers where each row represents a news text and each column represents a sentiment score. In the second step, we use this table to help us understand financial outcomes like stock returns or market trends. Using LLMs in the initial phase improves text representations compared to existing dictionary methods. This improvement comes from the extensive parameter space of LLMs and their training on a diverse range of language samples (Devlin et al., 2019). By enabling the use of LLMs, a wealth of knowledge becomes accessible for financial research (Huang et al., 2023). Our research primarily focuses on demonstrating the usefulness of LLM representations in modelling stock returns.

To achieve this, we conduct two separate econometric studies that leverage text mining to gain insights into the financial market. Firstly, we assess the relationship between financial news sentiment and stock returns using sentiment analysis techniques. This involves categorizing news sentiment based on the aggregated 3-day excess returns of stocks, providing a more dynamic and precise reflection of market response to news. Secondly, we conduct a regression analysis to determine how effectively LLM-derived sentiment scores predict stock returns on the following day. We utilize linear regression models with firm-specific and date-specific fixed effects, enabling a detailed evaluation of the predictive accuracy of LLMs, including OPT, BERT, and FinBERT, against traditional dictionary models. Finally, we explore practical applications through the development of distinct trading strategies based on LLM-derived sentiment scores. These strategies encompass creating long, short, and long-short portfolios, guided by sentiment scores from OPT, BERT, FinBERT and the Loughran-McDonald dictionary models. Our methodological design accounts for real-world trading conditions by incorporating transaction costs and aligning trade execution with news release timings. This comprehensive approach not only highlights the nuanced sentiment analysis capabilities of LLMs, but also demonstrates their practical value in formulating effective trading strategies in the financial market.

This investigation both complements and extends the scope of existing scholarly works that combine text processing and machine learning approaches to tackle a range of questions in financial research (Jegadeesh and Wu, 2013; Garcia, 2013; Hoberg and Phillips, 2016; Manela and Moreira, 2017; Hansen et al., 2018; Ke et al., 2020). Additionally, it contributes to the ongoing research focused on the correlation between news sentiment and collective stock market performance (Tetlock, 2007; Campbell et al., 2014; Baker et al., 2016; Calomiris and Mamaysky, 2019; Frankel et al., 2022). A unique aspect of our study lies in exploring the capabilities of LLMs. We hypothesize that these sophisticated LLMs have the potential to unearth more profound, and perhaps previously unrecognized, insights from textual data, leading to enhanced accuracy in predicting stock market reactions.

The paper is organized as follows: Section 2 describes the data and methods we used for our study. Section 3 reports the study's results and discusses them. Section 4 contains the conclusions.

## 2. Data and methods

### 2.1. Data

In our research, we primarily use two datasets: one from the Center for Research in Security Prices (CRSP) that includes daily stock returns, and another from Refinitiv with global news. The news data from Refinitiv comprises detailed articles and quick alerts, focusing on companies based in the U.S. The CRSP data provides daily return information for companies that trade on major U.S. stock exchanges. It includes details like stock prices, trading volumes and market capitalisation. We use this data to analyse the link between stock market returns and sentiment scores derived from LLMs.

Our analysis includes companies from the American Stock Exchange (AMEX), National Association of Securities Dealers Automated Quotations (NASDAQ) and New York Stock Exchange (NYSE) that appear in at least one news article. We apply filters to ensure the quality of our data. We only consider news articles related to individual stocks with available three-day returns. Moreover, we avoid redundancy by using a novelty score based on the similarity between articles: if a new article is too similar to an older article published within the past 20 days according to a cosine similarity score of 0.8 or more, we exclude it. This approach helps us to focus on unique information significant for our analysis.

Our study covers the period from January 1, 2010, to June 30, 2023. We matched 2,732,845 news about 6214 unique companies. After applying our filters, we were left with 965,375 articles. Our sample dataset is summarized in Table 1.

**Table 1**
Summary statistics of our U.S. news articles dataset after we applied filters. "All news" is the total count of news items from Refinitiv. "News for single stock" is the post-filtering count for articles exclusively associated with individual stocks. "Unique news" is the remaining count after excluding articles with a high degree of similarity (cosine similarity score higher than 0.8) to any other content published within the prior five business days, ensuring a dataset without redundant information.

| All news  | News for single stock | Unique news |
|-----------|----------------------|-------------|
| 2,732,845 | 1,865,372            | 965,375     |

Table 2 presents descriptive statistics of our dataset. We find that the mean daily return is 0.37% with a standard deviation of 0.18%. The sentiment scores derived from the OPT, BERT, and FinBERT models show a normal distribution around the median of 0.5, with slight variations in mean and standard deviation. In contrast, the Loughran-McDonald dictionary score exhibits a more positively skewed distribution with a mean of 0.68 and a higher standard deviation of 0.32, indicating a tendency towards more positive sentiment scores in our dataset.

**Table 2**
Descriptive statistics for daily stock returns expressed in percentage points and sentiment scores derived from the OPT, BERT, and FinBERT models, as well as the Loughran-McDonald (LM) dictionary. The table provides the mean, standard deviation (StdDev), minimum, median, maximum, and the total count of observations $N$ for each variable related to stock returns and sentiment scores from the mentioned models and dictionary.

|                    | Mean | StdDev | Minimum | Median | Maximum | $N$     |
|--------------------|------|--------|---------|--------|---------|---------|
| Daily return (%)   | 0.37 | 0.18   | −64.97  | −0.02  | 237.11  | 965,375 |
| BERT score         | 0.48 | 0.25   | 0       | 0.5    | 1       | 965,375 |
| OPT score          | 0.53 | 0.24   | 0       | 0.5    | 1       | 965,375 |
| FinBERT score      | 0.51 | 0.24   | 0       | 0.5    | 1       | 965,375 |
| LM dictionary score| 0.68 | 0.32   | 0       | 0.5    | 1       | 965,375 |

### 2.2. Methods

This study commences with the fine-tuning of pre-trained language models, specifically BERT and OPT, sourced from Hugging Face, to tailor their capabilities for specialized financial analysis (Hugging Face, 2023). LLMs, originally designed for broad linguistic comprehension, require significant adaptation to perform niche tasks such as producing a sentiment index through textual analysis of financial news with the ultimate aim of forecasting stock returns. This necessity enforces the adaptation phase, where the models are recalibrated after their original training on extensive data to prepare them for specific analytical functions (Radford et al., 2018).

In addition to the OPT and BERT LLMs, our analysis includes FinBERT, a variant of BERT pre-trained specifically for financial texts, and the Loughran and McDonald dictionary. Notably, FinBERT and the Loughran and McDonald dictionary do not necessitate the fine-tuning process, as they are already tailored for financial text analysis. FinBERT leverages BERT's architecture but is fine-tuned on financial texts, providing nuanced understanding in this domain (Huang et al., 2023). The Loughran and McDonald dictionary, a specialized lexicon for financial texts, aids in traditional textual analysis without the complexity of machine learning models (Loughran and McDonald, 2022).

Guided by the methodologies introduced by Alain and Bengio (2016), our approach adopts a probing technique, which is a form of feature extraction. This method builds on the models' pre-existing parameters, harnessing them to create features pertinent to text data, thereby facilitating the downstream task of sentiment analysis. To enhance the precision of our LLMs, we adapted and modified the methodology proposed by Ke et al. (2020). In our methodology, the process of fine-tuning the pre-trained OPT and BERT language models involves a specific focus on the aggregated 3-day excess return associated with each stock. This excess return is calculated from the day a news article is first published and extends over the two subsequent days. To elaborate, excess return is defined as the difference between the return of a particular stock and the overall market return on the same day. This calculation is not limited to the day the news is published; instead, it aggregates the returns for the following two days as well, providing a comprehensive three-day outlook.

Sentiment labels are assigned to each news article based on the sign of this aggregated three-day excess return. A positive aggregated excess return leads to a sentiment label of '1', indicating a positive sentiment. Conversely, a non-positive aggregated excess return results in a sentiment label of '0', suggesting a negative sentiment. Our approach of using a 3-day aggregated excess return for sentiment labelling plays a crucial role in refining our analysis. It follows the common practice in economics and finance of studying events that span multiple days (MacKinlay, 1997). This approach entails evaluating returns spanning from the day of the article's publication through the two following days. This technique is particularly beneficial in understanding the relationship between the sentiment in financial news and the corresponding movements in stock prices. We allocated 20% of the data randomly for testing and, from the remaining data pool, allocated another 20% randomly for validation purposes, resulting in a training set of 193,070 articles.

After completing the language model fine-tuning, our analysis continues with an empirical evaluation of these models in the context of U.S. financial news sentiment. A subset of 20% of these articles was set aside as a test sample, allowing for an unbiased evaluation of the models' predictive accuracy. Our analysis focused on the ability of OPT, BERT, FinBERT, and the Loughran-McDonald dictionary to accurately forecast the direction of stock returns based on news sentiment, particularly over a three-day period post-publication. To assess the models' performance, we calculated these statistical measures: accuracy, precision, recall, specificity, and the F1 score. Accuracy is the proportion of true results (both true positives and true negatives) among the total number of cases examined. Precision (positive predictive value) is the proportion of positive identifications that were actually correct. Recall (sensitivity) is the proportion of actual positives that were identified correctly. Specificity is the proportion of actual negatives that were correctly identified. The F1 score is the harmonic mean of precision and recall, providing a single metric that balances both quantities.

We subsequently conducted a regression analysis with the objective of investigating the influence of language model scores on the subsequent day's stock returns. The regression is modelled as

$$r_{i,n+1} = a_i + b_n + \gamma \cdot \mathbf{x}_{i,n} + \epsilon_{i,n}, \tag{1}$$

where $r_{i,n+1}$ is the return of stock $i$ on the subsequent trading day $n + 1$, $\mathbf{x}_{i,n}$ is a vector of scores from language models, and $a_i$ and $b_n$ are the fixed effects for firm and date. We employ double clustering for standard errors by firm and date, addressing potential concerns related to heteroscedasticity and autocorrelation. This regression framework facilitates an in-depth comparison of the predictive efficacy of different LLMs, including OPT, BERT, FinBERT and Loughran-McDonald dictionary variants with respect to stock returns.

Our choice of the linear regression model corresponds to a standard panel regression approach where article features $x_{i,n}$ are directly translated into the expected return $E(r_{i,n+1})$ of the corresponding stock on the next day. The simplicity of linear regression is chosen to emphasize the importance of text-based representations in financial analysis. By using linear models, we can focus on the impact of these representations without the added complexity of nonlinear modelling. This approach highlights the direct influence of textual data on financial predictions, ensuring a clear understanding of the role and effectiveness of text-based features in financial sentiment analysis.

Following our predictive analysis, our study extends to assess practical outcomes through the implementation of distinct trading strategies utilizing sentiment scores derived from OPT, BERT, FinBERT and the Loughran-McDonald dictionary models. To comprehensively evaluate these strategies, we construct various portfolios with a specific focus on market value-weighted approaches. For each language model, we create three types of portfolios: long, short, and long-short. The composition of these portfolios is contingent on the sentiment scores assigned to individual stocks every day. Specifically, the long portfolios comprise stocks with the highest 20% sentiment scores, while the short portfolios consist of stocks with the lowest 20% sentiment scores. Moreover, the long-short portfolios are self-financing strategies that simultaneously involve taking long positions in stocks with the highest 20% sentiment scores and short positions in stocks with the lowest 20% sentiment scores. We observe cumulative returns of these trading strategies considering transaction costs. We dynamically update these market value-weighted sentiment portfolios on a daily basis in response to changes in sentiment scores. This means that each day, we reevaluate and adjust the portfolios by considering the latest sentiment data. By doing so, we aim to capture the most current market conditions and enhance the effectiveness of our trading strategies.

This method allows us to test the real-world application of sentiment analysis findings without the influence of overall market movements. We base our stock choices on their market value, giving preference to larger, more stable companies, as these often represent safer, more reliable investments, and help reduce trading costs. We synchronize our trading decisions with the timing of news releases. For news reported before 6 am, we initiate trades at the market opening on that day, exploiting immediate reaction opportunities and close the position at the same date. For news appearing between 6 am and 4 pm, we initiate a trade with closing prices of the same day and exit the trade the next trading day. Any news coming in after 4 pm was used for trades at the start of the next trading day, adapting to market operating hours. To make our simulation more aligned with actual trading conditions, we included a transaction cost of 10 basis points for each trade, accounting for the typical costs traders would encounter in the market.

## 3. Results

### 3.1. Sentiment analysis accuracy in U.S. financial news

In this study, we used LLMs to analyse sentiment in U.S. financial news. We processed a dataset of 965,375 articles from Refinitiv, spanning from January 1, 2010, to June 30, 2023. We used 20% of these articles as a test set. We measured the accuracy of each model in predicting the direction of stock returns based on news sentiment. This accuracy indicates how well the model links the sentiment in financial news with stock returns over a three-day period. We evaluated four models: OPT, BERT, FinBERT and the Loughran-McDonald dictionary. Their performance in sentiment analysis is shown in Table 3.

**Table 3**
Language model performance metrics: accuracy, precision, recall, specificity, and the F1 score for each model. The OPT model is the most accurate, followed closely by BERT and FinBERT.

| Metric | OPT | BERT | FinBERT | Loughran-McDonald |
| --- | --- | --- | --- | --- |
| Accuracy | 0.744 | 0.725 | 0.722 | 0.501 |
| Precision | 0.732 | 0.711 | 0.708 | 0.505 |
| Recall | 0.781 | 0.761 | 0.755 | 0.513 |
| Specificity | 0.711 | 0.693 | 0.685 | 0.522 |
| F1 score | 0.754 | 0.734 | 0.731 | 0.508 |

The results show that the OPT model is the most accurate, followed closely by BERT and FinBERT. The Loughran-McDonald dictionary, a traditional finance text analysis tool, has significantly lower accuracy. This indicates that language models like OPT, BERT, and FinBERT are better at understanding and analysing complex financial news. The precision and recall values further support the superiority of the OPT model; its F1 score, which combines precision and recall, also confirms its effectiveness in sentiment analysis. These findings confirm that language models, particularly OPT, are valuable tools for analysing financial news and predicting stock market trends.

### 3.2. Predicting returns with LLM scores

This section assesses the ability of various LLMs to predict stock returns for the next day using regression models. Our regression with Eq. (1) uses LLM-generated scores from news headlines as the main predictors. To account for unobserved variations, these regressions include fixed effects for both firms and time, and we cluster standard errors by date and firm for added robustness. Table 4 provides our regression findings, focusing on how stock returns correlate with predictive scores from advanced LLMs, specifically OPT, BERT, FinBERT and the Loughran-McDonald dictionary models.

Our findings reveal the predictive capabilities of the advanced LLMs. The OPT model, in particular, demonstrates a strong correlation with next-day stock returns, as indicated by significant coefficients in different model specifications. The FinBERT model follows closely, showcasing its own robust predictive power. BERT scores, while more modest in their predictive strength, still show a statistically significant relationship with stock returns. We also observe that the predictive strength increases when both LLMs are used as independent variables in the same regression. In contrast, the Loughran-McDonald dictionary model exhibits the least predictive power among the models examined.

Our analysis suggests that several factors contribute to explain the different performance among OPT, BERT and FinBERT, notably model design, parameter scale, and the specificity of training data. OPT's expanded parameter space, exceeding that of BERT and FinBERT, alongside its advanced training methodologies, is likely to cause its superior forecasting accuracy in stock returns and portfolio management. Furthermore, the nuanced performance of FinBERT, despite its financial domain specialization, raises intriguing considerations. Our exploration detailed in Section 3.3 posits that the broader pre-training data diversity of BERT and the potential for overfitting in highly specialized models such as FinBERT might explain this unexpected outcome. These insights collectively emphasize the intricate balance between model specificity, scale, and training regimen in optimizing predictive performance within financial sentiment analysis.

The robustness of our regression models is further underscored by the inclusion of a substantial number of observations, ensuring a comprehensive and representative analysis. Additionally, the adjusted *R*-squared values, while moderate, indicate a reasonable level of explanatory power within the models. The reported AIC and BIC values aid in assessing model fit and complexity, further enriching our comparative analysis across different LLMs.

**Table 4**
Regression results of stock returns on LLM sentiment scores done with Eq. (1), which includes firm and time fixed effects (FE) represented by $a_i$ and $b_n$. The independent variable $x_{i,n}$ includes prediction scores from the language models. This analysis compares scores from OPT, BERT, FinBERT and Loughran-McDonald dictionary models, providing insights into their predictive abilities for stock market movements based on news sentiment. This analysis encompasses all U.S. common stocks with at least one news headline about the firm. *T*-statistics are presented in parentheses. Regressions 1 and 2 include two scores, regressions 3–6 only one. * $p < 0.05$. ** $p < 0.01$. *** $p < 0.001$.

| Regression | 1 | 2 | 3 | 4 | 5 | 6 |
|---|---|---|---|---|---|---|
| OPT score | 0.274*** | | 0.254*** | | | |
|  | (5.367) | | (4.871) | | | |
| BERT score | 0.142** | 0.091* | | 0.129* | | |
|  | (2.632) | (1.971) | | (2.334) | | |
| FinBERT score | | 0.257*** | | | 0.181*** | |
|  | | (5.121) | | | (4.674) | |
| LM dictionary score | | | | | | 0.083 |
|  | | | | | | (1.871) |
| Observations | 965,375 | 965,375 | 965,375 | 965,375 | 965,375 | 965,375 |
| R2 | 0.221 | 0.217 | 0.195 | 0.145 | 0.174 | 0.087 |
| R2 adjusted | 0.183 | 0.184 | 0.195 | 0.145 | 0.174 | 0.087 |
| R2 within | 0.021 | 0.022 | 0.017 | 0.009 | 0.016 | 0.002 |
| R2 within adjusted | 0.020 | 0.020 | 0.017 | 0.009 | 0.016 | 0.002 |
| AIC | 64,378 | 77,884 | 62,345 | 97,473 | 67,345 | 135,783 |
| BIC | 117,231 | 132,212 | 115,655 | 114,746 | 109,272 | 123,382 |
| RMSE | 5.32 | 11.12 | 4.21 | 14.12 | 9.75 | 23.54 |
| FE: date | X | X | X | X | X | X |
| FE: firm | X | X | X | X | X | X |

### 3.3. Performance of sentiment-based portfolios

Next, we assess the effectiveness of sentiment analysis in portfolio management by constructing various sentiment-based portfolios, including market value-weighted portfolios. These portfolios are developed using sentiment scores derived from different language models: OPT, BERT, FinBERT, and the Loughran-McDonald dictionary model. The investment strategies employed in our analysis can be described as follows: each LLM is utilised to create three distinct portfolios, one composed of stocks with top 20 percentile positive sentiment scores (long), another comprising stocks with top 20 percentile negative sentiment scores (short), and a self-financing long-short portfolio (L-S) based on both top 20 percentile negative and positive scores. Additionally, we include benchmark comparisons with value-weighted and equal-weighted market portfolios without considering sentiment scores. Value-weighted portfolios distribute investments based on the market capitalisation of each stock, while equal-weighted portfolios allocate investments equally to all stocks, regardless of market capitalisation. The selection of value-weighted and equal-weighted market portfolios was made to align with passive trading strategies, a widely acknowledged method in financial research (Fama and French, 1993; Carhart, 1997). We evaluate these strategies using key financial metrics, including the Sharpe ratio, mean daily returns, standard deviation of daily returns, and maximum drawdown.

The long-short OPT strategy demonstrates the most robust risk-adjusted performance, as evidenced by its superior Sharpe ratio indicated in Table 5. The Loughran-McDonald dictionary model-based strategy (L-S LM dictionary) laggs behind, particularly when compared to the value-weighted market portfolio. This highlights the varying effectiveness of different sentiment analysis models in guiding investment decisions and underscores the significance of model selection in sentiment-based trading.

**Table 5**
Descriptive statistics of trading strategies. The table presents the Sharpe ratio, mean daily return (MDR), daily standard deviation (StdDev) and the maximum daily drawdown (MDD) for the trading strategies based on the sentiment analysis models OPT, BERT, FinBERT, and Loughran-McDonald dictionary (LM dictionary), each comprising long (L), short (S), and long-short (L-S) portfolios. The portfolios are value-weighted for comparison to a value-weighted (VW) market portfolio, which is provided for benchmarking, as well as an equal-weighted (EW) portfolio. LM dictionary refers to a sentiment analysis approach that uses a dictionary of finance-specific terms developed by Loughran and McDonald.

|  | OPT | | | BERT | | | FinBERT | | |
|---|---|---|---|---|---|---|---|---|---|
|  | Long | Short | L-S | Long | Short | L-S | Long | Short | L-S |
| Sharpe ratio | 1.81 | 1.42 | 3.05 | 1.59 | 1.28 | 2.11 | 1.51 | 1.19 | 2.07 |
| MDR (%) | 0.32 | 0.25 | 0.55 | 0.25 | 0.21 | 0.45 | 0.22 | 0.18 | 0.39 |
| StdDev (%) | 2.18 | 2.91 | 2.49 | 2.49 | 3.19 | 2.68 | 2.59 | 3.31 | 2.81 |
| MDD (%) | −14.76 | −24.69 | −18.57 | −17.89 | −27.95 | −21.95 | −19.71 | −29.94 | −23.82 |
|  | LM dictionary | | | EW | | | VW | | |
|  | Long | Short | L-S | Long | Short | L-S | Long | Short | L-S |
| Sharpe ratio | 0.87 | 0.66 | 1.23 | 1.25 | 1.05 | 1.40 | 1.28 | 1.08 | 1.45 |
| MDR (%) | 0.12 | 0.13 | 0.22 | 0.18 | 0.15 | 0.33 | 0.19 | 0.16 | 0.35 |
| StdDev (%) | 3.54 | 4.13 | 3.74 | 2.90 | 3.70 | 3.20 | 2.95 | 3.75 | 3.25 |
| MDD (%) | −35.47 | −45.39 | −38.29 | −31.13 | −42.21 | −32.87 | −28.76 | −38.95 | −31.87 |

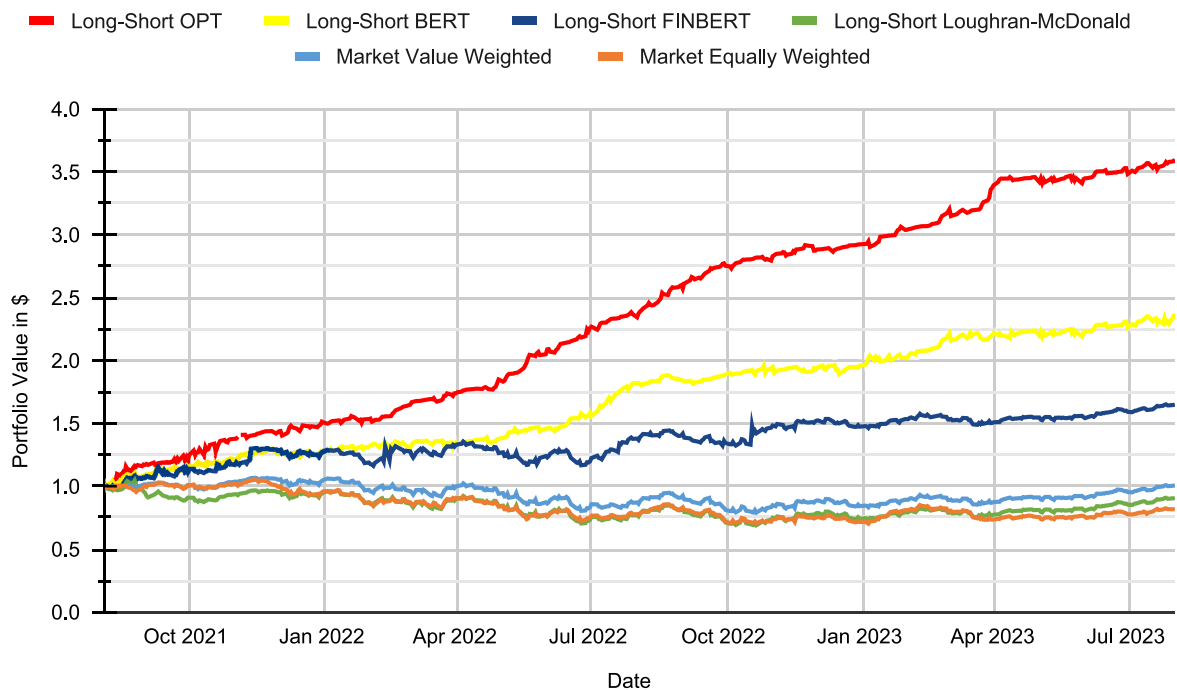

**Fig. 1.** Cumulative returns from investing $1 with value-weighted, zero-cost long-short portfolios based on OPT (red), BERT (yellow), FinBERT (dark blue) and the Loughran-McDonald dictionary (green), rebalanced daily with a 10 bps transaction cost. For comparison, we also show a value-weighted market portfolio (light blue) and an equal-weighted market portfolio (orange), both without transaction costs.

Finally, we examine the outcomes of trading strategies based on news sentiment including a 10 bps trading cost from August 2021 to July 2023. Fig. 1 illustrates the performance of various strategies, notably highlighting the long-short OPT strategy with an impressive 355% gain. This underscores the powerful predictive capability of advanced language models in forecasting market movements. Other strategies, such as long-short BERT and long-short FinBERT, also register significant gains of 235% and 165%, in stark contrast to traditional market portfolios, which barely exceed 1%. Conversely, the Loughran-McDonald dictionary model, extensively employed in finance research, managed only a 0.91% return. This pronounced disparity suggests that dictionary-based models do not effectively interpret the nuanced sentiments present in contemporary financial news as efficiently as more advanced language models. This analysis substantiates the importance of employing sophisticated language models in developing investment strategies based on news sentiment.

## 4. Conclusion

Our study has far-reaching implications for the financial industry, offering insights that could reshape market prediction and investment decision-making methodologies. By demonstrating the application of OPT and BERT models, we enhance the understanding of LLM applications in financial economics. This encourages further research into integrating artificial intelligence and LLMs in financial markets.

Notably, the advanced capabilities of LLMs surpass traditional sentiment analysis methods in predicting and explaining stock returns. We compared the performance of OPT, BERT and FinBERT scores to sentiment scores derived from conventional methods such as the sentiment score provided by the Loughran-McDonald dictionary model. Our analysis reveals that the latter basic model exhibits limited stock forecasting capabilities, with little to no significant positive correlation between their sentiment scores and subsequent stock returns. In contrast, complex models like OPT demonstrate the highest prediction power. For instance, a self-financing strategy based on OPT scores, buying stocks with positive scores and selling stocks with negative scores after news announcements, achieves a remarkable Sharpe ratio of 3.05 over our sample period, compared to a Sharpe ratio of 1.23 for the strategy based on the dictionary model.

The implications of our research reach beyond the financial industry to inform regulators and policymakers. Our research enhances our knowledge of the advantages and risks linked to the increasing use of LLMs in financial economics. As LLM usage expands, it becomes crucial to focus on their impact on market behaviour, information dissemination, and price formation. Our results add insights to the dialogue surrounding regulatory policies that oversee the use of AI in finance, thereby aiding in the establishment of optimal practices for incorporating LLMs into the operations of financial markets.

Our research offers tangible benefits to asset managers and institutional investors, presenting empirical data that demonstrates the strengths of LLMs in forecasting stock market trends. In our analysis covering August 2021 to July 2023, we observed that strategies utilizing news sentiment with advanced language models, notably the long-short OPT strategy, achieved remarkable gains, with an impressive 355% return. This starkly contrasts with the modest performance of traditional market portfolios and the Loughran-McDonald dictionary model, which only managed a 0.91% return. These findings highlight the significant advantage of employing sophisticated language models in developing effective investment strategies, marking a pivotal shift away from traditional sentiment analysis methods. Such evidence enables these professionals to make more informed choices regarding the integration of LLMs into their investment strategies. This could not only improve their performance but also decrease their dependence on traditional methods of analysis.

Our study contributes to the debate about the role of AI in finance, particularly through our investigation into how well LLMs predict stock market returns. By investigating both the possibilities and the boundaries of LLMs in the domain of financial economics, we open the way for further research aimed at creating more advanced LLMs specifically designed for the distinctive needs of the finance sector. Our goal in highlighting the potential roles of LLMs in financial economics is to foster ongoing research and innovation in the field of finance that is driven by artificial intelligence.

**CRediT authorship contribution statement**

**Kemal Kirtac:** Writing – original draft, Conceptualization, Methodology, Data Curation, Software, Investigation, Formal analysis.
**Guido Germano:** Writing — review & editing, Validation, Supervision.

**Declaration of competing interest**

The authors declare that they have no known competing financial interests or personal relationships that could have appeared to influence the work reported in this paper.

**Data availability**

The authors do not have permission to share data.